\documentclass[preprint,showpacs,preprintnumbers,amsmath,amssymb]{revtex4}

\usepackage{graphicx}
\usepackage{dcolumn}
\usepackage{bm}

\usepackage{color}

\begin{document}


\title{Non-monotonic Dynamics in Frustrated Ising Model with Time-Dependent Transverse Field}

\author{Shu Tanaka}
\thanks{
Present address: Research Center for Quantum Computing,
Interdisciplinary Graduate School of Science and Engineering,
Kinki University,
3-4-1 Kowakae, Higashi-Osaka, Osaka 577-8502, Japan.
}
\email{shu-t@alice.math.kindai.ac.jp}
\affiliation{%
Institute for Solid State Physics, University of Tokyo,
5-1-5, Kashiwanoha, Kashiwa-shi, Chiba 277-8581, Japan.
}
\author{Seiji Miyashita}
\email{miya@spin.phys.s.u-tokyo.ac.jp}
\affiliation{
Department of Physics, University of Tokyo, 7-3-1, Hongo, Bunkyo-ku,
Tokyo, 113-0033, Japan.\\
CREST, JST, 4-1-8 Honcho, Kawaguchi, Saitama 332-0012, Japan
}%

\date{\today}

\begin{abstract}
We study how the degree of ordering depends on the strength of the 
thermal and quantum fluctuations in frustrated systems by investigating 
the correlation function of the order parameter. Concretely, we compare 
the equilibrium spin correlation function in a frustrated lattice which 
exhibits a non-monotonic temperature dependence (reentrant type 
dependence) with that in the ground state as a function of the 
transverse field that causes the quantum fluctuation. We find the 
correlation function in the ground state also shows a non-monotonic 
dependence on the strength of the transverse field. 
We also study the real-time dynamics of the spin correlation function under 
a time-dependent field. After sudden decrease of the temperature, we found 
non-monotonic changes of the correlation function reflecting the static 
temperature dependence, which indicates that an effective temperature of 
the system changes gradually. For the quantum system, we study the 
dependence of changes of the correlation function on the sweeping speed 
of the transverse field. Contrary to the classical case, the correlation 
function varies little in a rapid change of the field, though it shows a 
non-monotonic change when we sweep the field slowly.
\end{abstract}

\maketitle

\section{Introduction}

Static and dynamic properties of frustrated systems have been studied extensively in the past several decades~\cite{Toulouse-1977,Liebmann-book-1986,Kawamura-1998,Diep-book-2005}.
Many model materials have been developed and theoretical studies of frustrated systems have been of increasing significance~\cite{Kageyama-1999,Bramwell-2001,Nakatsuji-2005,Ishii-2009}. 
In frustrated systems, there are highly degenerated ground states and thermal and/or quantum fluctuations generate a peculiar density of states.
Frustration causes various interesting phenomena such as the so-called ``order by disorder'' and ``reentrant phase transition'', etc. Fluctuations prevent the system from ordering in unfrustrated systems, however, in some frustrated systems, fluctuations stabilize some ordered structures due to a kind of entropy effect.
The ordering phenomena due to fluctuations are called ``order by disorder''~\cite{Villain-1980,Henley-1989,shu-t-2007d,Kamiya-2009,Tomita-2009}. There have been various examples of the order by disorder phenomena and also reentrant phase transitions~\cite{Nakano-1968,Syozi-1968,Syozi-1972,Fradkin-1976,Miyashita-1983,Kitatani-1985,Kitatani-1986,Azaria-1987,Miyashita-2001,shu-t-2005}.

Frustration plays an important role in dynamic properties as well as static properties. Since there are many degenerate states in frustrated systems, relaxation processes of physical quantities often show characteristic features. It is well known that slow relaxation appears in random systems such as spin glasses~\cite{Mezard-book-1987,Fisher-book-1993,Young-book-1998,Vincent-2006} and diluted Ising models~\cite{Bray-1988,Nowak-1989,Kawasaki-1993}. 
We also found that slow relaxation processes take place also in non-random systems. We have pointed out that frustration causes a stabilization of the spin state by a kind of screening effect, and a very slow dynamics appears. We recently proposed a mechanism of slow relaxation in a system without an energy barrier for the domain wall~\cite{shu-t-2009f}.

Recently, we have also studied the relation between the temperature dependence of static quantities and their dynamics after a sudden change of the temperature in some frustrated systems where the temperature dependence of the spin correlation function is non-monotonic. In that study, we found that the non-monotonic relaxation of the correlation function indicates a picture of relaxation of an effective temperature of the system~\cite{shu-t-2007c}.

Quantum fluctuations also cause some ordering structure in frustrated systems as well as thermal fluctuations~\cite{Moessner-2000}. The so-called quantum dimer model is the simplest model that exhibits the order by quantum disorder. This system has a rich phase diagram as a function of the on-site potential and the kinetic energy. When the on-site potential and the kinetic energy are equal, the model corresponds to a Rokhsar-Kivelson point~\cite{Rokhsar-1988}, while the model corresponds to an Ising antiferromagnet with the transverse field on its dual lattice, when the on-site potential is zero. Recently, ``order by quantum disorder'' was also studied in a transverse Ising model on a fully frustrated lattice from a viewpoint of adiabatic quantum annealing~\cite{Matsuda-2009}.

The nature of the dynamics in quantum systems with time-dependent external fields has also been studied extensively~\cite{Landau-1932,Zener-1932,Stueckelberg-1932,Rosen-1932,Chiorescu-2003}. It is an important issue to investigate the real-time dynamics in quantum system for control of the quantum state by external fields. Recently, quantum annealing or quantum adiabatic evolution have been studied from a viewpoint of quantum dynamics~\cite{Kadowaki-1998,Farhi-2001,Santoro-2002,Das-book-2005}. They are the methods to obtain the ground state in complex systems. Now, the quantum annealing method is adopted on a wide scale, for example, clustering problem and Bayes inference which are important topics in information science and information engineering~\cite{shu-t-2009c,shu-t-2009d}.

The purpose of the present study is to understand the similarity and difference between the effects of the thermal fluctuations and the quantum fluctuations in frustrated systems. In this paper, we study the dynamic properties of a frustrated Ising spin system with a time-dependent transverse field. In Section 2, we introduce a frustrated Ising system with decorated bonds, and we review equilibrium and dynamical properties of this classical system. In Section 3, we study the ground-state properties of this model with a transverse field, and we investigate the dynamical nature of this model with a time-dependent transverse field. We also consider the microscopic mechanism of the non-monotonic dynamics of the spin correlation function. In Section 4, we summarize the present study.

\section{Classical Decorated Bond System}

\subsection{Equilibrium Properties}
\begin{figure}[b]
 \begin{center}
  \includegraphics[scale=0.75]{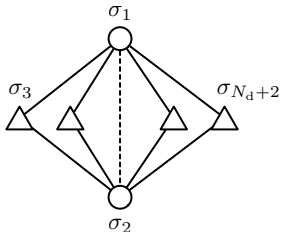}
  \caption{
Decorated bond system. The solid and the dotted lines represent ferromagnetic interactions $-J$ and an antiferromagnetic interaction $J_0$, respectively. The circles and the triangles denote the system spins, $\sigma_1$ and $\sigma_2$, and the decoration spins $\sigma_i$ ($i = 3, \cdots, N_{\rm d}+2$), respectively. This figure shows the case of $N_{\rm d} = 4$.
  }
  \label{fig:model}
 \end{center}
\end{figure}

\begin{figure}[b]
 \begin{center}
  \includegraphics[trim=40 0 40 0,scale=0.5]{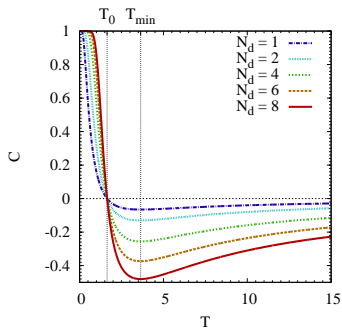}
  \caption{
  (Color online) The system correlation function $C$ given by Eq.~(\ref{Keff}) between the system spins $\sigma_1$ and $\sigma_2$ as a function of the temperature. In this case, $T_{\rm min} \simeq 3.64$ and $T_0 \simeq 1.64$.
  }
  \label{graph:classical-cor-keff}
 \end{center}
\end{figure}

First we briefly review static properties of the model depicted in Fig.~\ref{fig:model}.  The circles and the triangles denote the system spins $\sigma_1$ and $\sigma_2$ and the decoration spins $\sigma_i$ ($i=3, \cdots N_{\rm d}+2$), respectively, where $N_{\rm d}$ is the number of the decoration spins.  Both of the system spins and the decoration spins are $S=1/2$ Ising spins.
The Hamiltonian of this system is given by
\begin{eqnarray}
 {\mathcal H}_{\rm c} = J_0 \sigma_1 \sigma_2
- J \sum_{i=3}^{N_{\rm d}+2} \left( \sigma_1 + \sigma_2 \right)
\sigma_i,
\end{eqnarray}
where $J_0$ and $J$ are positive. 
The direct bond $J_0$ tends to have an antiferromagnetic correlation while the interactions through the decoration spin $J$ tend to have ferromagnetic one. We will study effects of the competition between them. If we set $J_0= N_{\rm d}J$, the interactions cancel out each other at zero temperature. Because the interaction through the decoration spin is weakened at finite temperatures as we will see below, the spin correlation is antiferromagnetic at finite temperatures in this case.
In this paper we focus the case where the spin correlation function changes the sign at a temperature, and therefore we study the case $J_0 < N_{\rm d}J$.
Concretely, we adopt the case $J_0= N_{\rm d}J/2$. The consequences obtained in the present paper hold qualitatively for any choice of the ratio $J_0/N_{\rm d}J$ as long as it is less than unity. We take $J$ as the energy unit from now on. Figure \ref{graph:classical-cor-keff} shows the equilibrium spin correlation function between the system spins $\sigma_1$ and $\sigma_2$ (hereafter, we call it the ``system correlation function'' and denote it by $C$) as a function of the  temperature in the cases of $N_{\rm d} = 1,2,4,6$, and $8$.
In Fig.~\ref{graph:classical-cor-keff}, $T_{\rm min}$ and $T_0$ represent the temperatures where $C$ takes the minimum value and $C=0$, respectively. In the present case, $T_{\rm min} \simeq 3.64$ and $T_0 \simeq 1.64$.

The system correlation function $C$ behaves non-monotonically as a function of temperature.  This non-monotonic behavior comes from a kind of entropy effect.  At $T=0$, all the spins align in the same direction, because energetically it is the most favorable state.  At finite temperatures the decoration spins can flip due to the thermal fluctuations.  When each decoration spin has the values $\pm 1$ randomly, the ferromagnetic interactions depicted by the solid lines in Fig.~\ref{fig:model} are weakened, and the direct antiferromagnetic interaction becomes dominant.  The states, in which the decoration spins align randomly, are entropically favorable.  
Therefore, at a high temperature, $C$ becomes negative due to the entropy effect. This is why $C$ behaves non-monotonically as a function of the temperature.  
Following Refs.~\cite{Nakano-1968,Syozi-1968,Syozi-1972,Fradkin-1976,Miyashita-1983,Kitatani-1985,Kitatani-1986,Azaria-1987,Miyashita-2001,shu-t-2005}, the effective Hamiltonian ${\cal H}_{\rm eff} = - J_{\rm eff} \sigma_1 \sigma_2$ is defined by
\begin{eqnarray}
 \sum_{\sigma_3 = \pm 1} \cdots 
 \sum_{\sigma_{N_{\rm d}+2} = \pm 1} 
  {\rm e}^{-\beta \mathcal{H}} 
  = \left( 2 \cosh 2 \beta J \right)^{\frac{N_{\rm d}}{2}}{\rm e}^{-\beta \mathcal{H}_{\rm eff}}.
\end{eqnarray}
The reduced coupling of the system spins is obtained as
\begin{eqnarray}
 K_{\rm eff} = \beta J_{\rm eff} 
  = \frac{N_{\rm d}}{2} \log \cosh \left( 2 \beta J \right) 
  - \frac{N_{\rm d}}{2} \beta J.
\label{Keff}
\end{eqnarray}
The temperature dependence of $C$ can be calculated analytically by tracing out the degree of freedom of the decoration spins:
\begin{eqnarray}
\label{Keff}
C = \left\langle \sigma_1 \sigma_2 \right\rangle = \tanh K_{\rm eff}.
\end{eqnarray}
Here it should be noted that the effective interaction Eq.~(\ref{Keff}) is proportional to $N_{\rm d}$, and thus the temperature at which $C=0$ is the same for all $N_{\rm d}$ and that temperature of the minimum $C$ as well. 

%

\subsection{Kinetic dynamics}
\label{sec:classical-T-change}

In the previous section, we showed the equilibrium properties of the decorated bond system. In this subsection, we study relaxation processes of $C$ after the temperature is decreased suddenly. We adopt the Glauber Ising model for the time evolution:
\begin{eqnarray}
 \nonumber
 &&\frac{\partial P \left( \sigma_1, \cdots, \sigma_i, \cdots,
		   \sigma_{N_{\rm d} + 2}; t \right)}{\partial t} \\
 \nonumber
 &&= \sum_i P \left( \sigma_1, \cdots, -\sigma_i, \cdots, \sigma_{N_{\rm
	     d}+2}; t \right) w_{-\sigma_i \to \sigma_i}\\
 &&- \sum_i P \left( \sigma_1, \cdots, \sigma_i, \cdots, \sigma_{N_{\rm
	     d}+2}; t \right) w_{\sigma_i \to -\sigma_i},
\end{eqnarray}
where $ P \left( \sigma_1, \cdots, \sigma_i, \cdots, \sigma_{N_{\rm d} + 2}; t \right) \equiv P\left( \left\{ \sigma_i \right\};t\right)$ and $w_{\sigma_i \to -\sigma_i}$ represent the probability distribution at time $t$ and the transition probability from $\sigma_i$ to $-\sigma_i$, respectively.
The transition probability $w_{\sigma_i \to -\sigma_i}$ is given by
\begin{widetext}
\begin{eqnarray}
 w_{\sigma_i \to -\sigma_i} =
  \frac{P_{\rm eq}^{\left( T \right)}\left( \sigma_1, \cdots, -\sigma_i, \cdots,
		   \sigma_{N_{\rm d}+2}\right)}
  {P_{\rm eq}^{\left( T \right)}\left( \sigma_1, \cdots, \sigma_i, \cdots,
		   \sigma_{N_{\rm d}+2}\right) +
  P_{\rm eq}^{\left( T \right)}\left( \sigma_1, \cdots, -\sigma_i, \cdots,
		   \sigma_{N_{\rm d}+2}\right)},
\end{eqnarray}
\end{widetext}
where $P_{\rm eq}^{\left( T \right)}\left( \sigma_1, \cdots, \sigma_i, \cdots,
\sigma_{N_{\rm d}+2}\right) \equiv P_{\rm eq}^{\left( T \right)} \left( \left\{ \sigma_i \right\}\right)$ denotes the equilibrium probability distribution at the
temperature $T$.

In Ref.~\cite{shu-t-2007c}, we studied the dynamics of $C$ after a sudden change of the temperature from a finite temperature $T_1$ to a finite temperature $T_2$. 
In this paper, we change the temperature suddenly from $T_1 = \infty$ to a finite temperature $T_2$. 
The dynamics of $C$ is given by
\begin{eqnarray}
 C\left( t \right) = \sum_{\left\{ \sigma_i \right\}}
  P \left( \{\sigma_i\}; t \right)
  \sigma_1 \sigma_2,
\end{eqnarray}
where $P \left( \left\{ \sigma_i \right\}; t\right)$ is the probability of the  configuration $\left\{ \sigma_i \right\}$ at the time $t$. 

From the equilibrium properties, it is trivial that $C$ behaves non-monotonically, when we decrease the temperature slowly enough. 
However, it is not clear whether $C$ behaves monotonically or not when the temperature is changed suddenly.

\begin{figure}[t]
 \begin{center}
  $\begin{array}{ccc}
   \includegraphics[trim=40 0 40 0,scale=0.4]{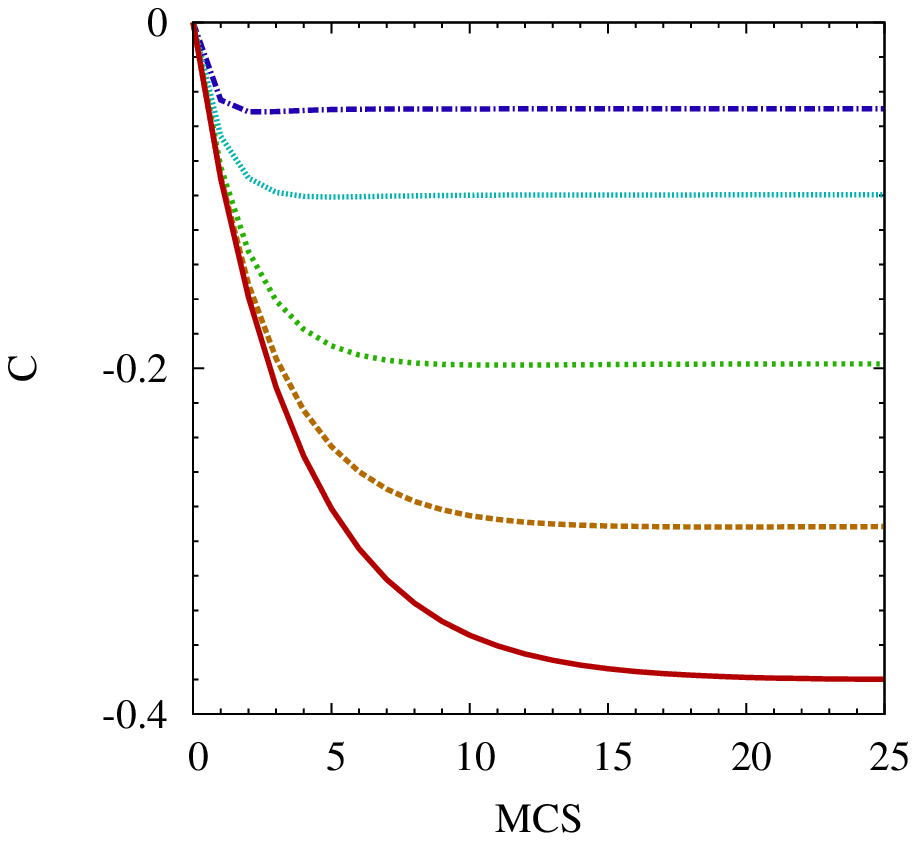}
    & \hspace{1mm} &
    \includegraphics[trim=40 0 40 0,scale=0.4]{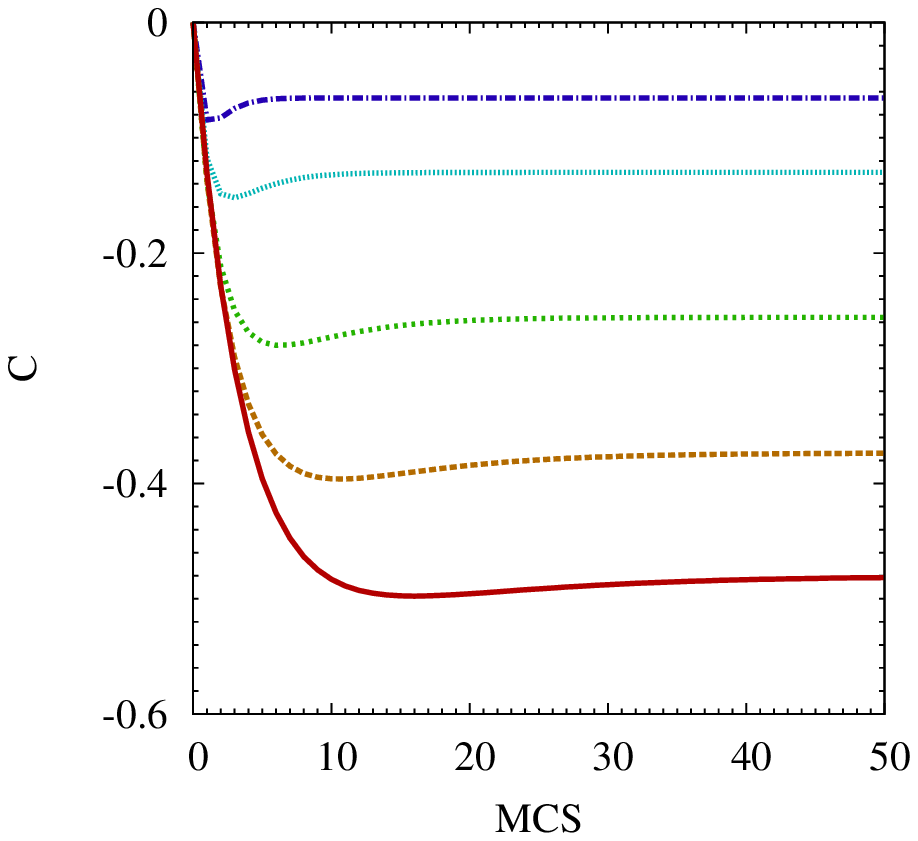}\\
   ({\rm a}) T_2=7.29 & \hspace{1mm} & ({\rm b}) T_2=3.64 (\simeq T_{\rm min})\\
    \includegraphics[trim=40 0 40 0,scale=0.4]{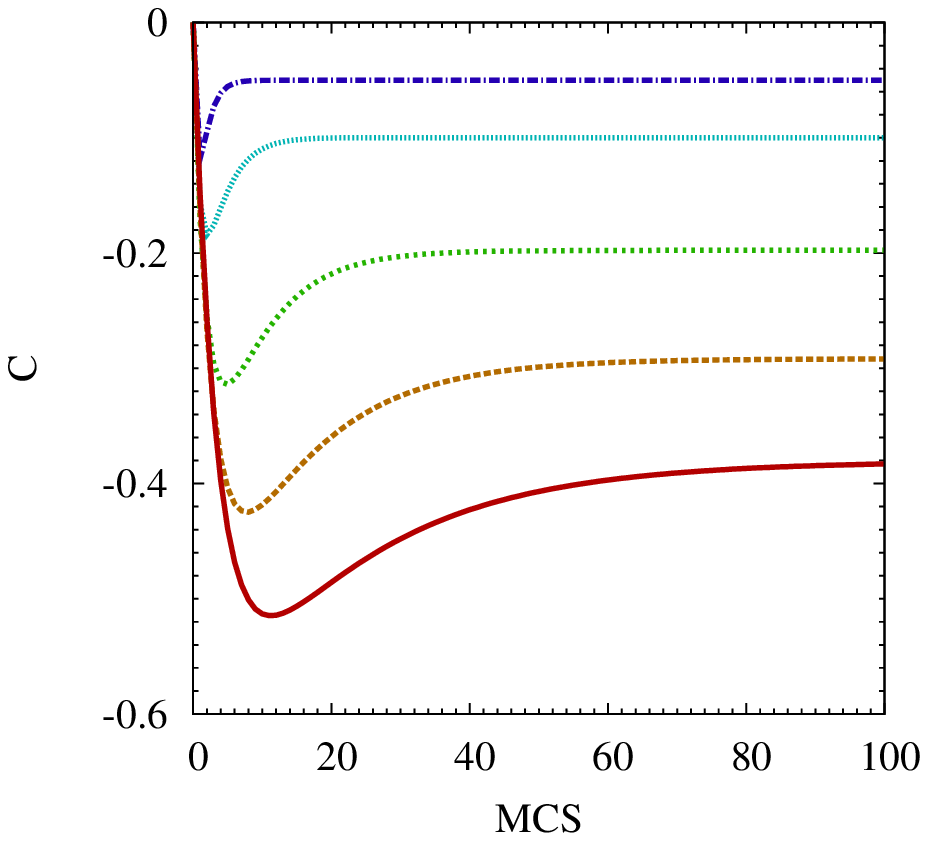}
& \hspace{1mm} &
\includegraphics[trim=40 0 40 0,scale=0.4]{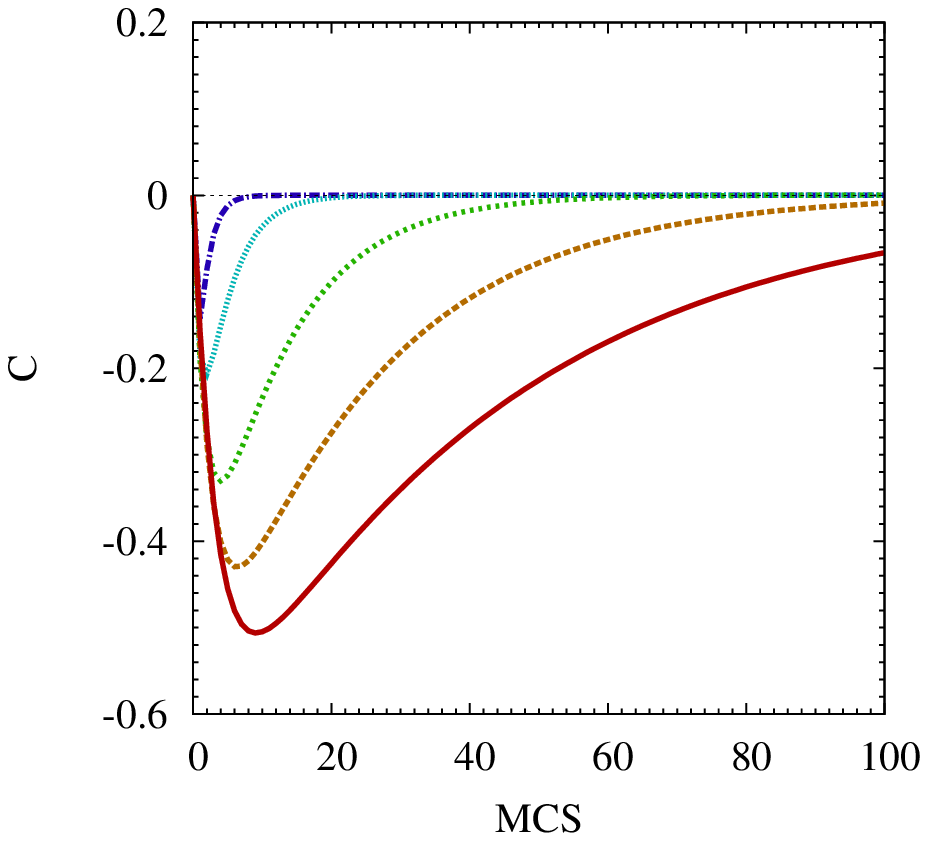}\\
     ({\rm c}) T_2=2.35 & \hspace{1mm} & ({\rm d}) T_2=1.64 (\simeq T_0)\\
      \includegraphics[trim=40 0 40 0,scale=0.4]{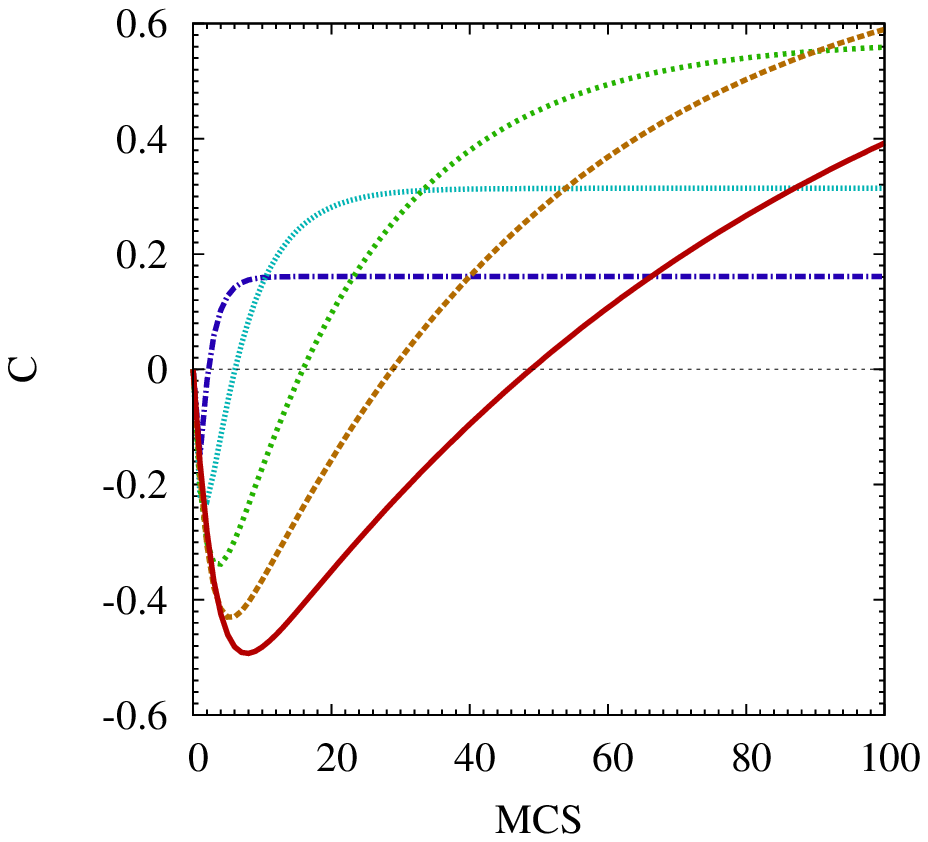}
     & \hspace{1mm} &
     \includegraphics[trim=40 0 40 0,scale=0.4]{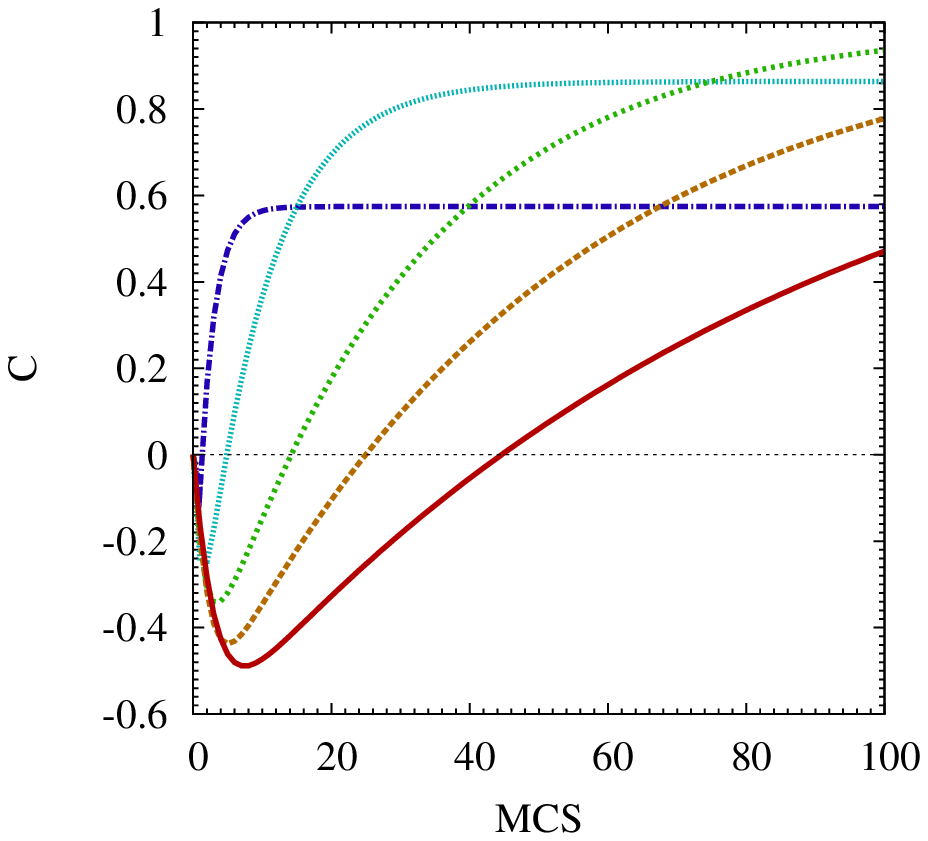}\\
 ({\rm e}) T_2=1 & \hspace{1mm} & ({\rm f}) T_2=0.5
   \end{array}$
  \caption{
  (Color online) Dynamics of $C$ for $T_2 = 7.29$, $3.64 (\simeq T_{\rm min})$, $2.35$, $1.64 (\simeq T_0)$, $1$, and $0.5$. The initial condition is set to be the equilibrium probability distribution at $T=\infty$.
The lines have the same interpretations as in Fig.~\ref{graph:classical-cor-keff}.
Note the different scales on the axes in the different parts.
  }
  \label{graph:glauber-dynamics}
 \end{center}
\end{figure}

The initial condition is set in the equilibrium state at $T_1=\infty$, in other words, the initial condition is the uniform distribution: $P\left( \left\{ \sigma_i \right\}; t = 0 \right) = 1/2^{N_{\rm d}+2}$. We suddenly decrease the temperature to a finite temperature $T_2$. 
Figure \ref{graph:glauber-dynamics} shows the time evolutions of $C$ for $N_{\rm d}=2,4,6,8$, and $10$ in the cases of $T_2 = 7.29$, $3.64(\simeq T_{\rm min})$, $2.35$, $1.64(\simeq T_0)$, $1$, and $0.5$. As we saw in Fig.~\ref{graph:classical-cor-keff}, at $T_2 = 7.29$, $3.64$, and $2.35$, the equilibrium value of $C$ is negative. At $T_2 = 1.64$, the equilibrium value of $C$ is zero, and at $T_2 = 1.0$ and $0.5$, it is positive.

In the case of $T_2 = 7.29 > T_{\rm min}$, $C$ monotonically decreases toward the equilibrium value. On the other hand, non-monotonic relaxations appear in the cases of $T_2=3.64$, $2.35$, $1.64$, $1$, and $0.5$ which are lower than $T_{\rm min}$.
Such a nature was pointed out in Ref.~\cite{shu-t-2007c}.
Although the temperature $T$ is suddenly changed to a low temperature $T_2$, the property of the system, i.e., the correlation function, shows a gradual change which seems to follow the gradual decrease of the temperature. Therefore, we regard that a kind of ``effective temperature'' of the system decreases gradually.
This non-monotonic behavior takes place as long as $T_2 < T_{\rm min}$.
This fact indicates that such non-monotonic relaxation does not depend on the equilibrium value of $C$ at the final temperature $T=T_2$.
It should be noted that as $N_{\rm d}$ increases, the time evolution shows stronger non-monotonicity. 

\section{Decorated Bond System with Transverse Field}

In the previous section, we introduced the decorated Ising model and studied its equilibrium and dynamical properties. In this section, we study the ground-state properties of this model with a transverse field, and the dynamical nature of this model with a time-dependent transverse field.

\subsection{Ground-State Properties}

\begin{figure*}[t]
 \begin{center}
  $\begin{array}{ccccc}
  \includegraphics[trim=40 0 40 0,scale=0.4]{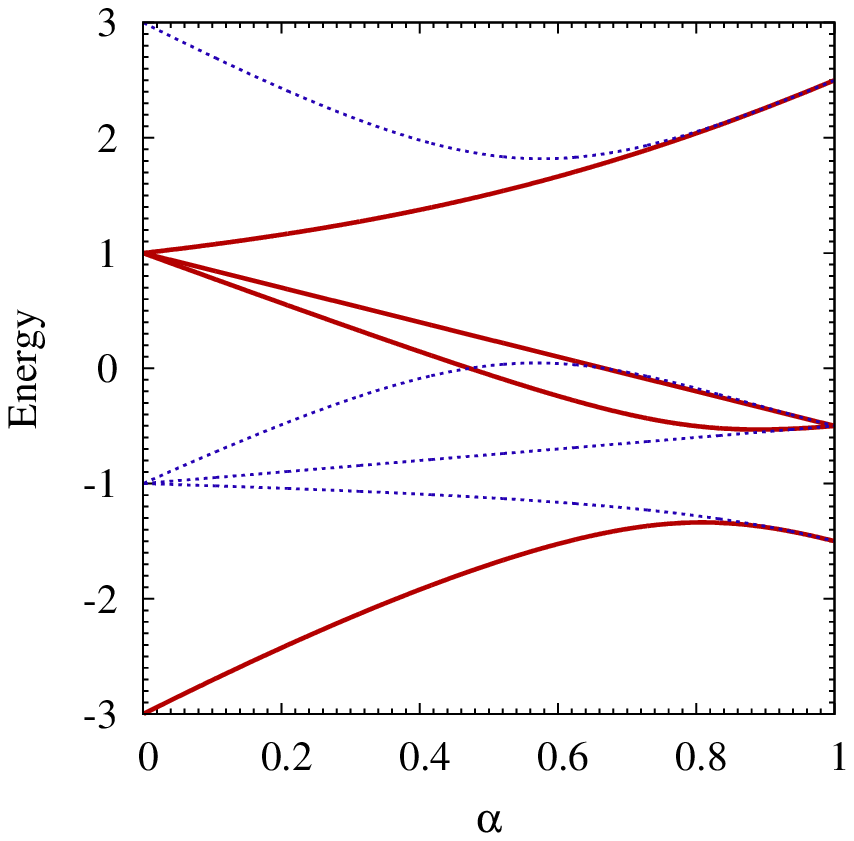}
    & \hspace{1mm} &
  \includegraphics[trim=40 0 40 0,scale=0.4]{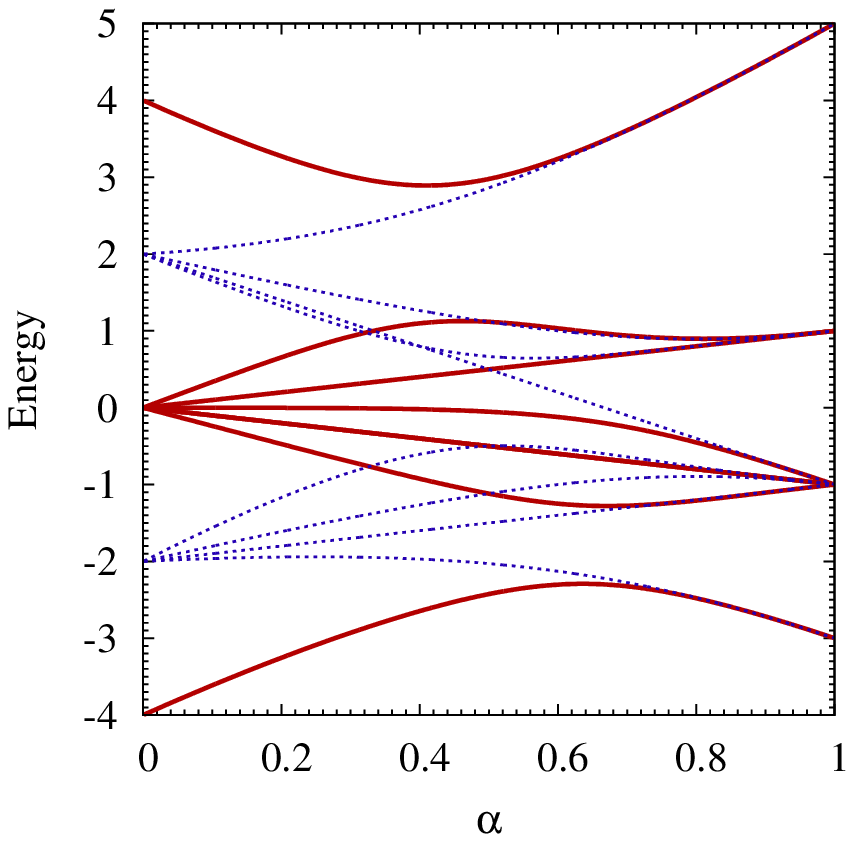}
    & \hspace{1mm} &
  \includegraphics[trim=40 0 40 0,scale=0.4]{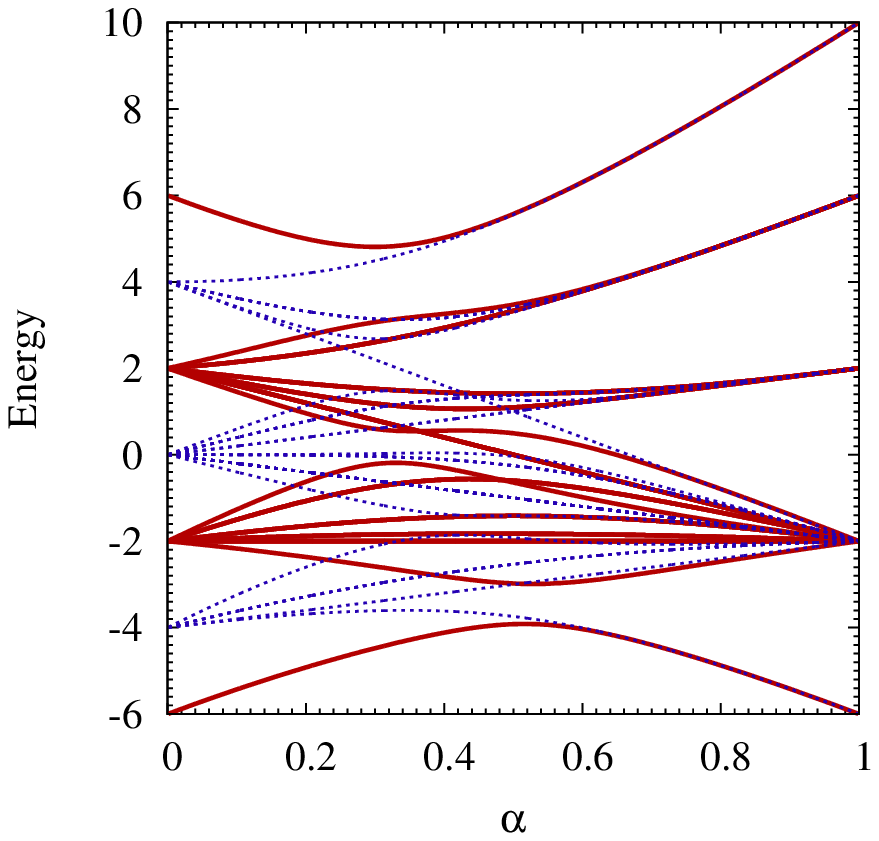}\\
    ({\rm a}) N_{\rm d} = 1&\hspace{1mm}& ({\rm b}) N_{\rm d} =
     2&\hspace{1mm}& ({\rm c}) N_{\rm d} = 4
   \end{array}$
  \caption{
  (Color online) Eigenenergies as functions of $\alpha$ for $N_{\rm d} = 1$, $2$, and $4$. Solid and dotted curves denote the eigenenergy for symmetric wavefunctions and for antisymmetric wavefunctions, respectively.
  }
  \label{graph:eng}
 \end{center}
\end{figure*}

The Hamiltonian is given by
\begin{equation}
 \label{Eq:quantum-totHam}
\mathcal{H}\left( \alpha \right) = \alpha {\cal H}_{\rm c} + \left( 1 - \alpha \right)
  {\cal H}_{\rm q} \,\,\,\,\,\,\, (0 \le \alpha \le 1), 
\end{equation}
where
\begin{equation}
\mathcal{H}_{\rm c} = \frac{N_{\rm d}}{2} J \sigma_1^z \sigma_2^z
- J \sum_{i=3}^{N_{\rm d}+2} \left( \sigma_1^z + \sigma_2^z \right)
\sigma_i^z, 
\end{equation}
and 
\begin{equation}
\mathcal{H}_{\rm q} = - \sum_{i=1}^{N_{\rm d}+2} \sigma_i^x.
\end{equation}
Here, $\mathcal{H}_{\rm c}$ and $\mathcal{H}_{\rm q}$ represent the classical part and the quantum part of the total Hamiltonian, respectively. 
The operators $\sigma_i^x$ and $\sigma_i^z$ are the Pauli matrices,
\begin{eqnarray}
 \sigma_i^x = 
  \left(
   \begin{array}{cc}
    0 & 1\\
    1 & 0
   \end{array}
  \right), 
  \,\,\,\,\,\,
  \sigma_i^z = 
  \left(
   \begin{array}{cc}
    1 & 0\\
    0 & -1
   \end{array}
  \right).
\end{eqnarray}
Hereafter, we call $\alpha$ the ``quantum parameter''. The total Hamiltonian (Eq.~(\ref{Eq:quantum-totHam})) with $\alpha = 1$ corresponds to the completely classical Hamiltonian. The total Hamiltonian possesses parity symmetry.
Now we make use of the all-spin-flip operator:
\begin{eqnarray}
 {\cal P} = \prod_{i=1}^{N_{\rm d}+2} \sigma_i^x
\end{eqnarray}
to study the parity symmetry. Total Hamiltonian and the all-spin-flip operator commute:
\begin{eqnarray}
 \left[ {\cal H}\left( \alpha \right), {\cal P} \right] = 0, \,\,\,\,\,\, \left( {\rm for} \,\,\forall \alpha\right)
\end{eqnarray}
Symmetric wave function and antisymmetric wave function are defined as follows:
\begin{eqnarray}
 &\left| \Phi_{\rm s} \right\rangle = 
  \sum_{\left\{ \sigma \right\}}' a_{\sigma}
  \left( \left| \sigma \right\rangle 
   + {\cal P} \left| \sigma \right\rangle \right),\\
 &\left| \Phi_{\rm as} \right\rangle = 
  \sum_{\left\{ \sigma \right\}}' a_{\sigma}
  \left( \left| \sigma \right\rangle 
   - {\cal P} \left| \sigma \right\rangle \right),
\end{eqnarray}
where $ \left| \sigma \right\rangle $ denote a state of a spin configuration, and
$\sum_{\left\{ \sigma \right\}}'$ denotes summation over all the spin configurations fixing $\sigma_1 = +1$. $\left| \Phi_{\rm s} \right\rangle$  and $\left| \Phi_{\rm as} \right\rangle$ represent symmetric and antisymmetric wave functions, respectively. Because ${\cal P} \left| \Phi_{\rm s} \right\rangle = \left| \Phi_{\rm s} \right\rangle$ and ${\cal P} \left| \Phi_{\rm as} \right\rangle =-\left| \Phi_{\rm as} \right\rangle$ are satisfied, the total Hamiltonian can be block-diagonalized according to the symmetry. 
Note that the ground state of ${\cal H}\left( \alpha \right)$ with arbitrary $\alpha$ is a symmetric wave function.  Figure \ref{graph:eng} shows the eigenenergies as functions of the quantum parameter $\alpha$ for $N_{\rm d} = 1$, $2$, and $4$.  The solid and the dotted curves in Fig.~\ref{graph:eng} denote the eigenenergy values for symmetric and antisymmetric wavefunctions, respectively.

\begin{figure}[b]
 \begin{center}
  \includegraphics[trim=40 0 40 0,scale=0.5]{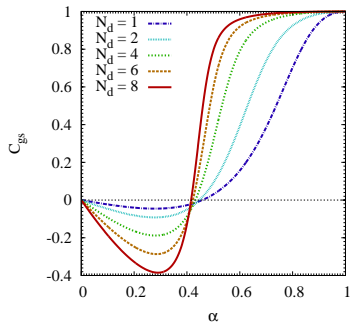}
  \caption{
  (Color online) The system correlation function of the ground state $C_{\rm gs}$ in the ground state as a function of
  the quantum parameter $\alpha$.
  }
  \label{graph:timdec-cor}
 \end{center}
\end{figure}

We consider the system correlation function in the ground state:
$C_{\rm gs}\left( \alpha \right) = \left\langle \Psi_{\rm gs}\left( \alpha \right) \right|
\sigma_1^z \sigma_2^z
\left| \Psi_{\rm gs}\left( \alpha \right) \right\rangle$,
where $\left| \Psi_{\rm gs}\left( \alpha \right) \right\rangle$ denotes the ground state of the Hamiltonian (Eq.~(\ref{Eq:quantum-totHam})) with the quantum parameter $\alpha$. Figure \ref{graph:timdec-cor} shows the system correlation function in the ground state as a function of the quantum parameter $\alpha$.
The system correlation function of the ground state behaves non-monotonically as well as the equilibrium value at the finite temperature as shown in Fig.~\ref{graph:classical-cor-keff}. This fact indicates a similarity between the thermal fluctuation and the quantum fluctuation, although some details are different, {\it e.g.} the points where the system correlation function $C_{\rm gs}$ becomes zero and the minimum values depend on the number of the decoration spins in the quantum case, whereas they do not depend on the number of the decoration spins in the classical case. 

\begin{figure}[t]
 \begin{center}
  \includegraphics[scale=0.6]{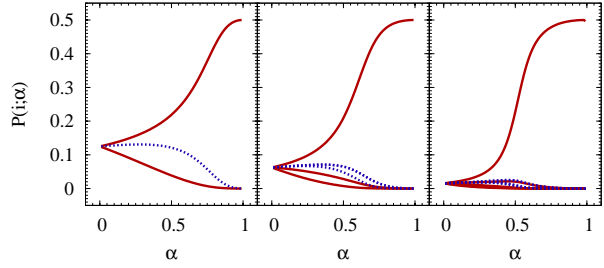}
  \caption{
  (Color online) Adiabatic limit of the probability distribution function 
$P\left( i;\alpha \right)$ for $N_{\rm d}=1,2,$ and $4$ (from left to right).
  The solid and the dotted lines denote the probability of the states where the system correlation functions are positive and negative, respectively.
Some of $P\left( i;\alpha \right)$s give the same curves and they are overlapped each other.
  }
  \label{graph:prob-dis-adia}
 \end{center}
\end{figure}

To consider the microscopic mechanism of this non-monotonic behavior, we calculate the probability of the classical bases
\begin{equation}
P \left( i;\alpha \right) =  \left| \left\langle i | 
	     \Phi_{\rm gs} \left( \alpha \right) \right\rangle \right|^2,
\end{equation}
where
$\left| i \right\rangle = 
\left| +,+,+,\cdots,+ \right\rangle$ and $\left| -,+,+,\cdots,+
\right\rangle$
etc.
Figure \ref{graph:prob-dis-adia} shows the probability distribution as a function of the quantum parameter.
The solid and the dotted lines in Fig.~\ref{graph:prob-dis-adia} denote $P\left( i;\alpha \right)$ for the states in which $\sigma_1^z \sigma_2^z = +1$, and  $\sigma_1^z \sigma_2^z = -1$ states, respectively.
In the case of $N_{\rm d}=1$, though there are eight classical configurations:
$\left( \sigma_1,\sigma_2,\sigma_3 \right)
=\left( +,+,+\right)$,
$\left( -,+,+\right)$,
$\left( +,-,+\right)$,
$\left( -,-,+\right)$,
$\left( +,+,-\right)$,
$\left( -,+,-\right)$,
$\left( +,-,-\right)$, and
$\left( -,-,-\right)$,
only three configurations 
$\left( +,+,+\right)$,
$\left( +,+,-\right)$, and
$\left( +,-,-\right)$
give different values because of the symmetry.
Thus we see only three lines in the left panel of Fig.~\ref{graph:prob-dis-adia}.
By the same reason, we see only five lines and eight lines in the middle and the right panels of Fig.~\ref{graph:prob-dis-adia}.

\subsection{Real-Time Dynamics}

In the previous section, we studied the ground-state properties of the decorated bond system with the transverse field. The system correlation function behaves non-monotonically as a function of the quantum parameter $\alpha$. Now, we consider the real-time dynamics of the system correlation function in the quantum case by the time-dependent Schr\"odinger equation.
The system correlation function $C(t)$ is defined by 
\begin{eqnarray}
 C\left( t \right) = \left\langle \psi\left( t \right) \right|
  \sigma_1^z \sigma_2^z
  \left| \psi \left( t \right)\right\rangle, 
\end{eqnarray}
where $\left| \psi \left( t \right)\right\rangle$ denotes the wavefunction at time $t$.
Now we consider the time-dependent Hamiltonian expressed by
\begin{eqnarray}
 \mathcal{H} \left( t \right)
  = \frac{t}{\tau} \mathcal{H}_{\rm c} +
  \left( 1 - \frac{t}{\tau} \right) \mathcal{H}_{\rm q},
\end{eqnarray}
where $\tau^{-1}$ represents the sweeping speed. Here, $t/\tau$ corresponds to $\alpha$ in the previous section.

The initial condition is set to be the ground state of ${\cal H}\left( t = 0\right) = {\cal H}_{\rm q}$ such as
\begin{eqnarray}
\left| \Psi\left( t=0\right) \right\rangle
= \left| \rightarrow,\cdots, \rightarrow
\right\rangle,
\end{eqnarray}
where $\left| \rightarrow \right\rangle = (\left| \uparrow \right\rangle
+ \left| \downarrow \right\rangle)/\sqrt{2}$.
Because this ground state is a symmetric wavefunction, the wavefunction is also symmetric after the sweeping.
If the sweeping speed is slow enough ({\it i.e.} the adiabatic limit), the system correlation function behaves as given in Fig.~\ref{graph:prob-dis-adia}.

\begin{figure}[t]
 \begin{center}
  $\begin{array}{ccc}
   \includegraphics[trim=40 0 40 0,scale=0.4]{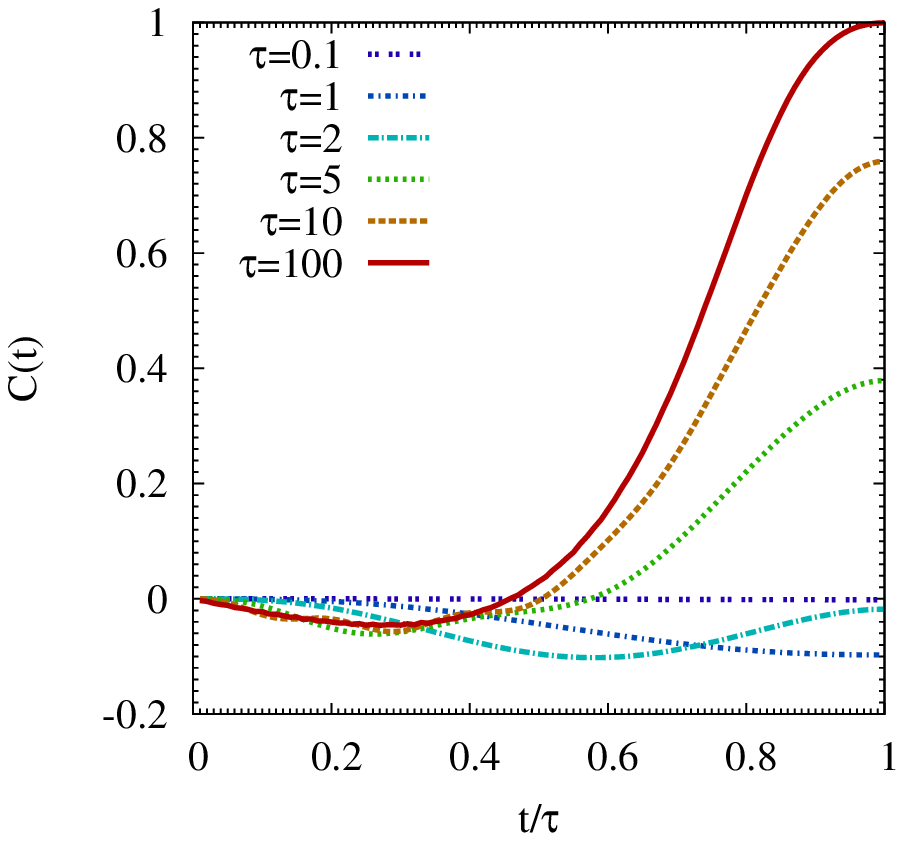}
    & \hspace{1mm} &
    \includegraphics[trim=40 0 40 0,scale=0.4]{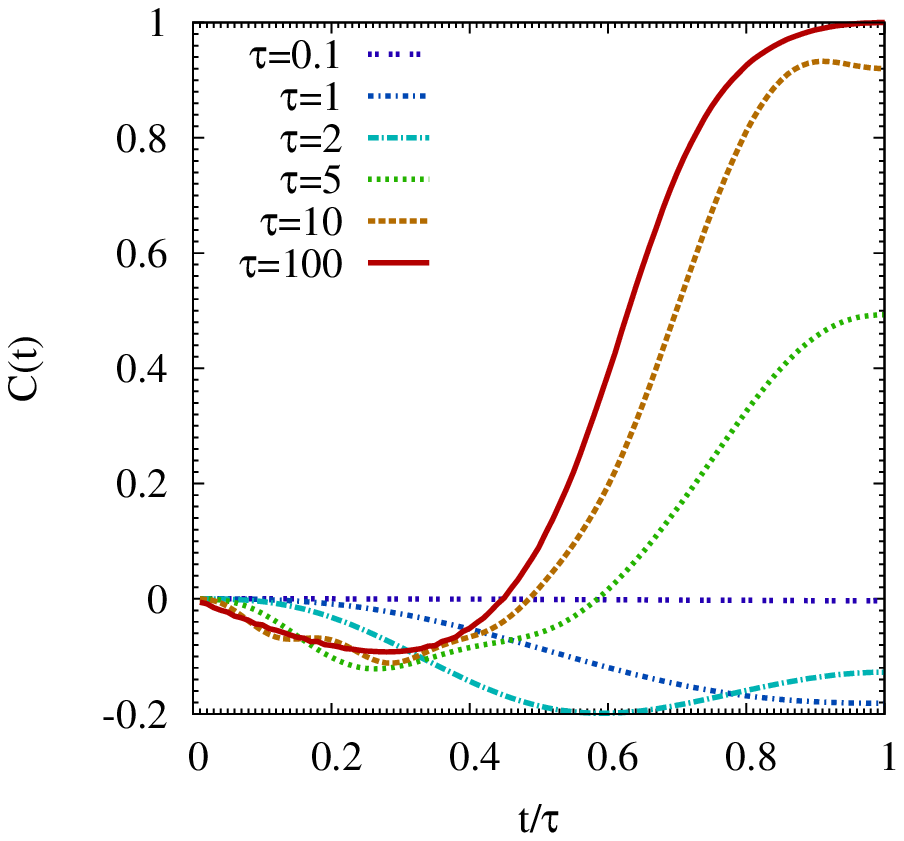}\\
    ({\rm a}) N_{\rm d} = 1& \hspace{1mm} & ({\rm b}) N_{\rm d} = 2 \\
    \includegraphics[trim=40 0 40 0,scale=0.4]{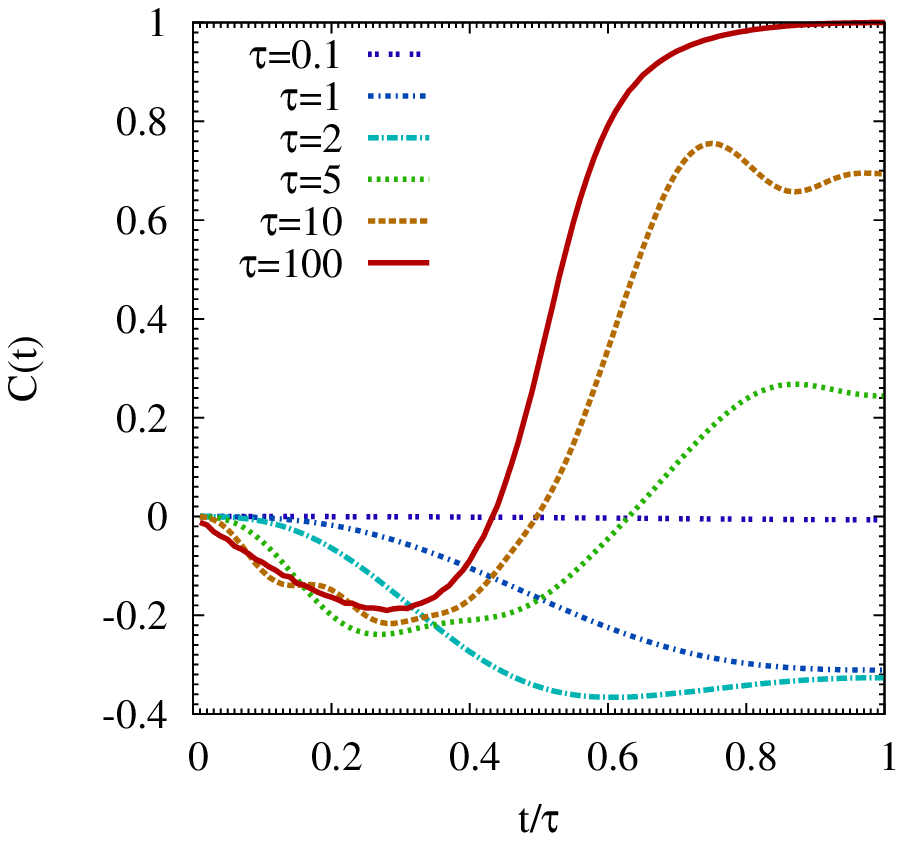}&
\hspace{1mm} & \includegraphics[trim=40 0 40 0,scale=0.4]{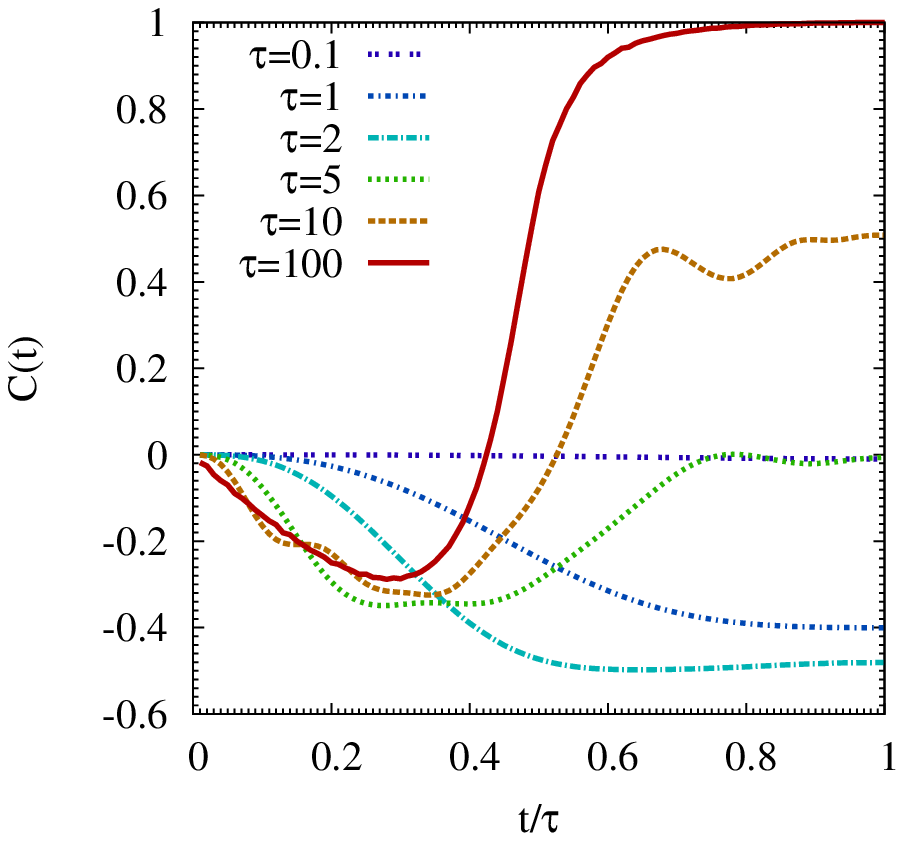}\\
({\rm c}) N_{\rm d} = 4& \hspace{1mm} & ({\rm d}) N_{\rm d} = 6 \\
     \includegraphics[trim=40 0 40 0,scale=0.4]{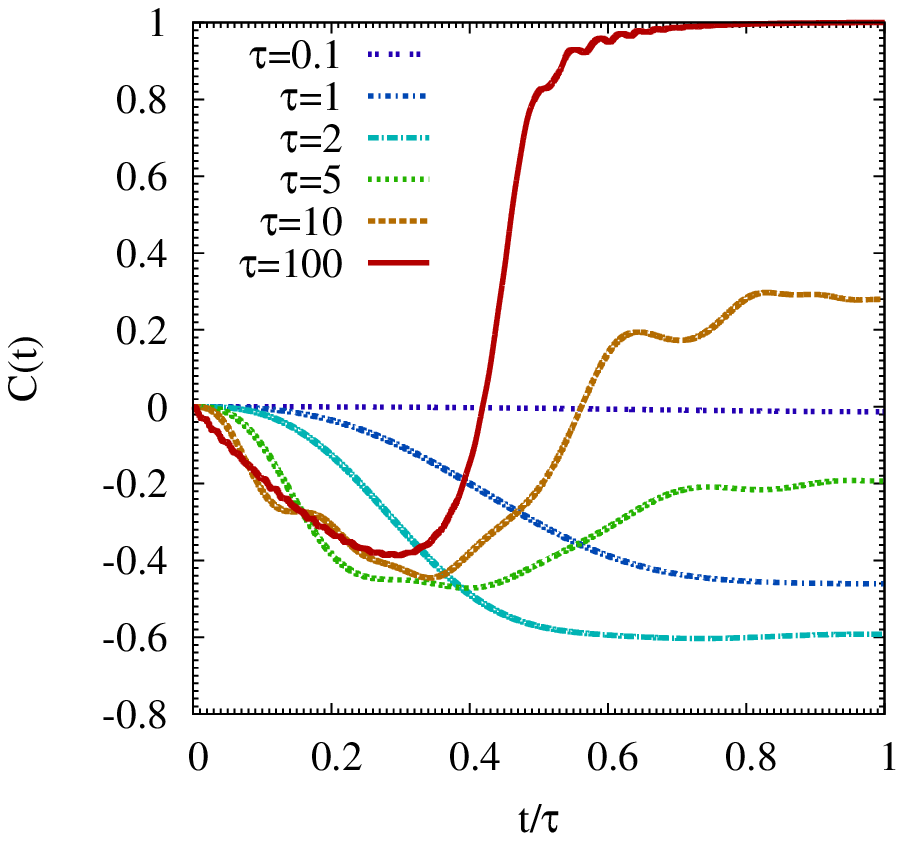}
     & \hspace{1mm} &
     \includegraphics[trim=40 0 40 0,scale=0.4]{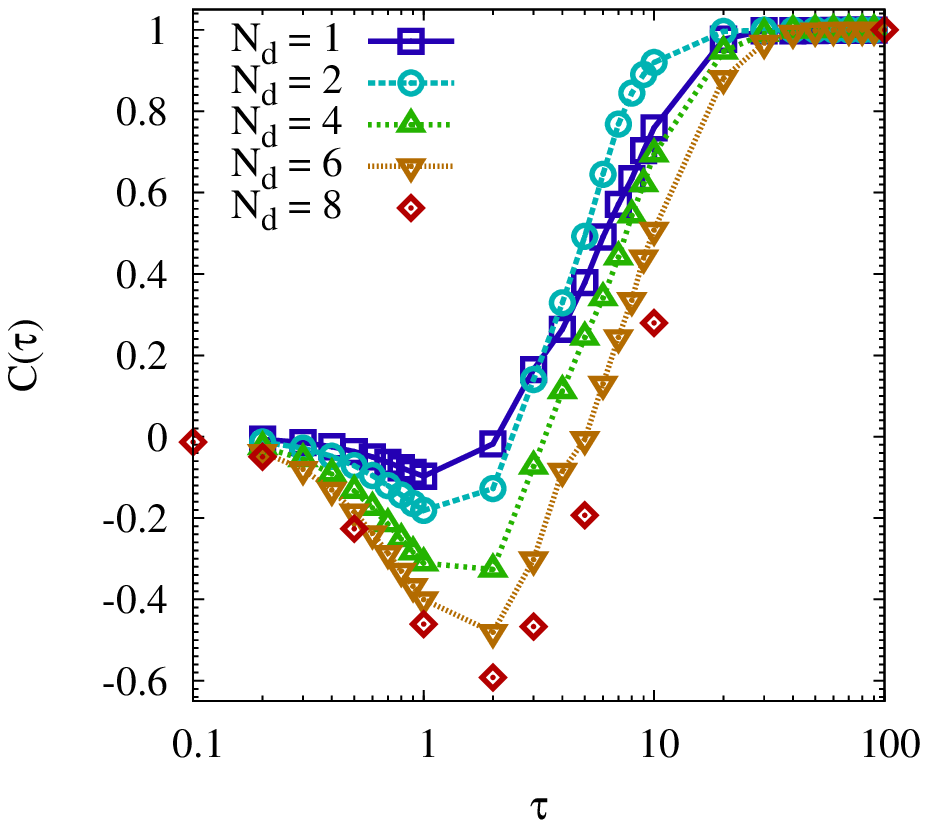}\\
    ({\rm e}) N_{\rm d} = 8 & \hspace{1mm} & ({\rm f})
   \end{array}$
  \caption{
  (Color online) Figures (a)-(e) give the real time dynamics of the system correlation function 
  for $N_{\rm d}=1$, $2$, $4$, $6$, and $8$, respectively.
  (f) The final values of the system correlation function as a function of $\tau$ (semilog scale).
  }
  \label{graph:dyn-12468}
 \end{center}
\end{figure}

Figure \ref{graph:dyn-12468} (a)-(e) shows the real-time dynamics of the system correlation function $C(t)$ in the cases of $N_{\rm d} = 1$, $2$, $4$, $6$, and $8$ for several values of $\tau$.
For large values of $\tau$, $C(t)$ is non-monotonic as a function of time, while it changes little when $\tau$ is small. 
The final value $C(\tau)$ is non-monotonic as a function of $\tau$. For the largest value of $\tau$, $C(t=\tau)\simeq 1$ which is the value in the ground, while it is nearly zero for the small value $\tau=0.1$. In between, $C(\tau)$ takes negative values. The change becomes significant when $N_{\rm d}$ increases.
We summarize these dependence in Fig.~\ref{graph:dyn-12468}(f).
Here we see that, for small values of $\tau$, $C(\tau)$ moves to a negative value monotonically as $\tau$ increases, and that
$C(\tau)$ has a minimum point at an intermediate value of $\tau$.

The negative $C(\tau)$ is considered to be attributed to the negative value of $C_{\rm gs}$ at an intermediate value of $\alpha$. And thus, we again find a gradual change of the effective temperature of the system as we saw in the case of temperature changing protocol of the classical system as stated in Section \ref{sec:classical-T-change}.
However, if we change $\alpha$ fast (small $\tau$), we find monotonic changes which are very different from the classical case.
This indicates a difference between thermal fluctuation and quantum fluctuation effect.

\begin{figure}[t]
 \begin{center}
  $\begin{array}{ccc}
   \includegraphics[trim=40 0 40 0,scale=0.4]{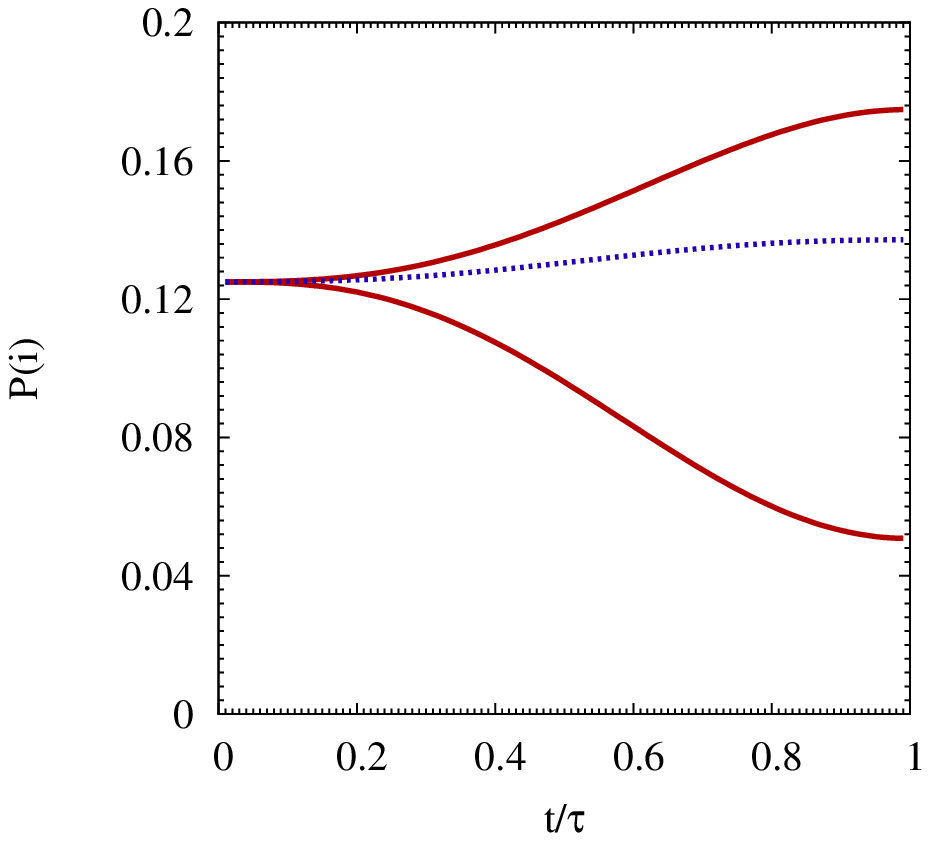}
   & \hspace{1mm} &
\includegraphics[trim=40 0 40 0,scale=0.4]{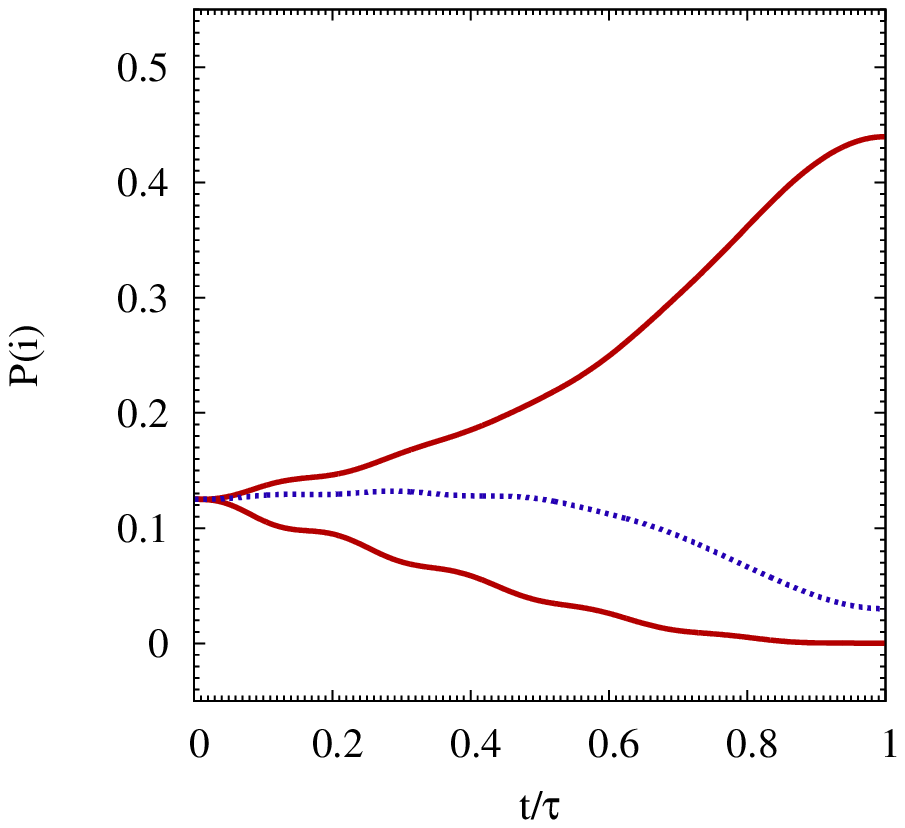}\\
({\rm a}) N_{\rm d} = 1, \tau = 1 &\hspace{1mm}& ({\rm b}) N_{\rm d} = 1, \tau = 10\\
   \includegraphics[trim=40 0 40 0,scale=0.4]{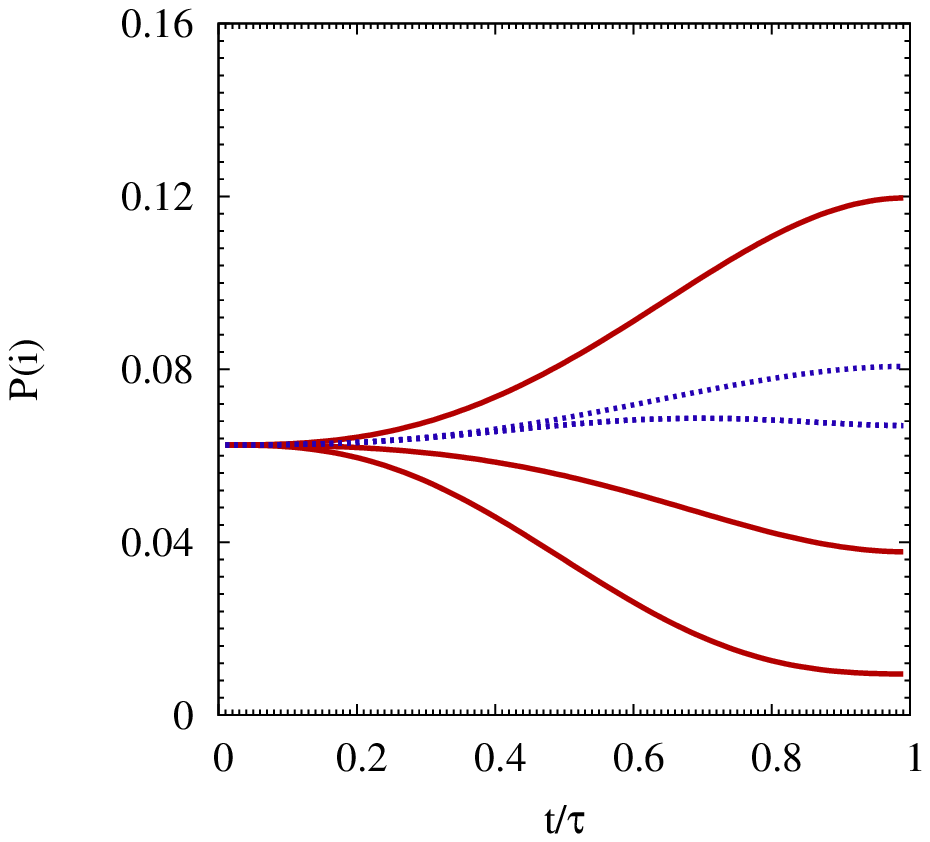}
   & \hspace{1mm} &
\includegraphics[trim=40 0 40 0,scale=0.4]{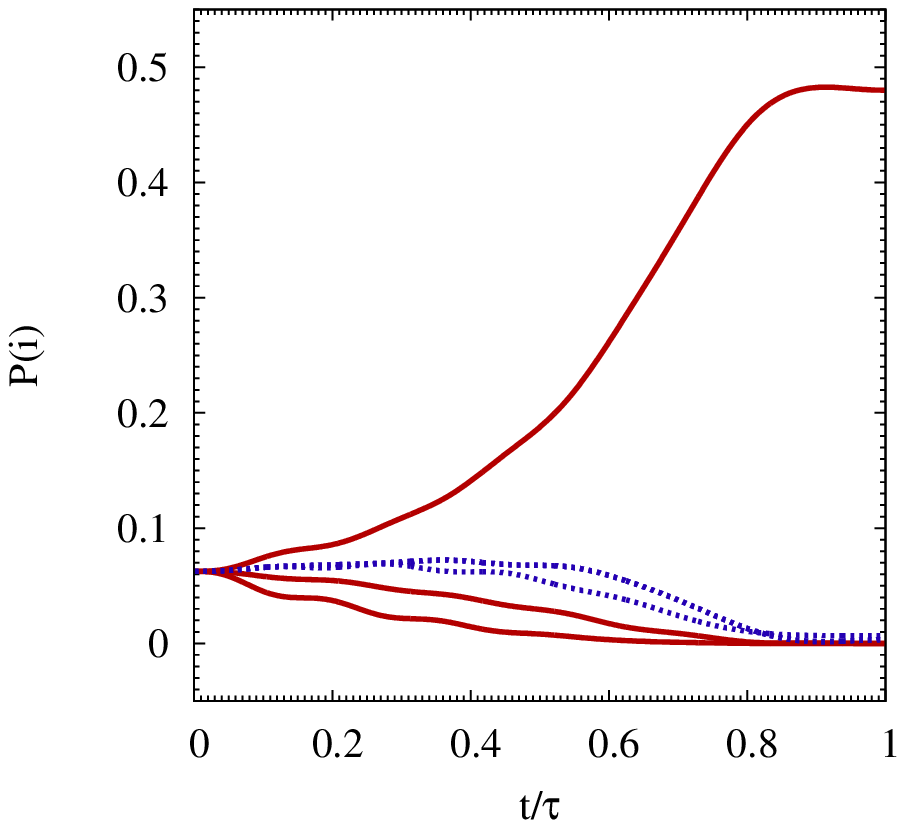}\\
({\rm c}) N_{\rm d} = 2, \tau = 1 &\hspace{1mm}& ({\rm d}) N_{\rm d} = 2, \tau = 10\\
   \includegraphics[trim=40 0 40 0,scale=0.4]{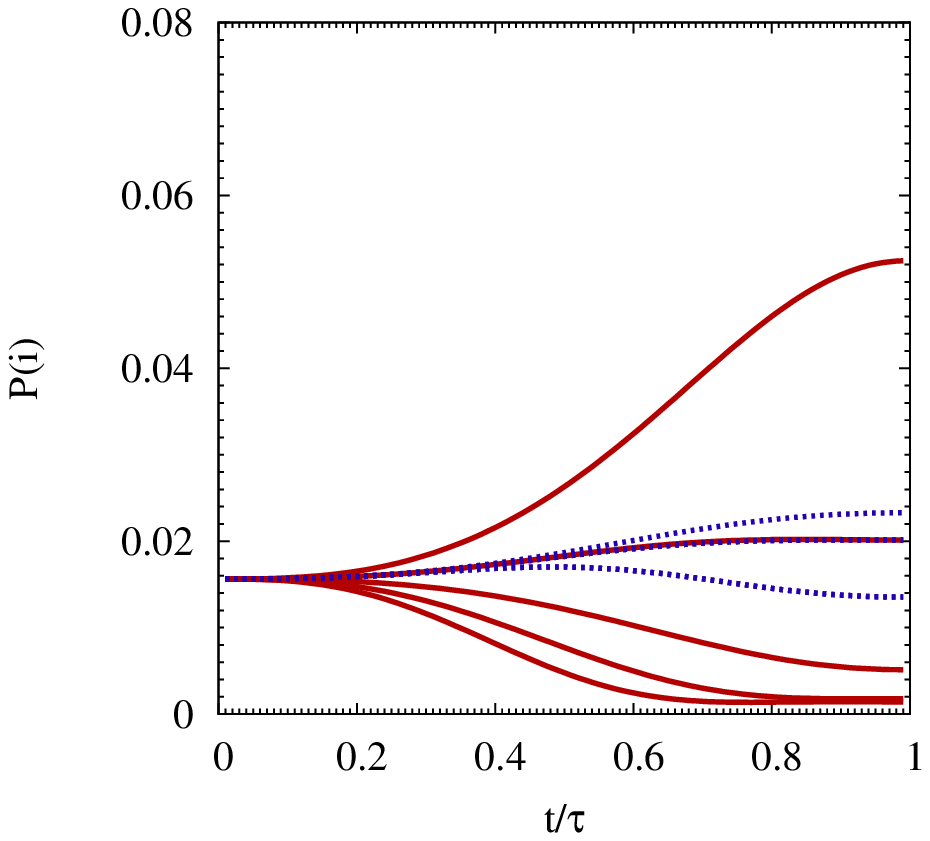}
   & \hspace{1mm} &
\includegraphics[trim=40 0 40 0,scale=0.4]{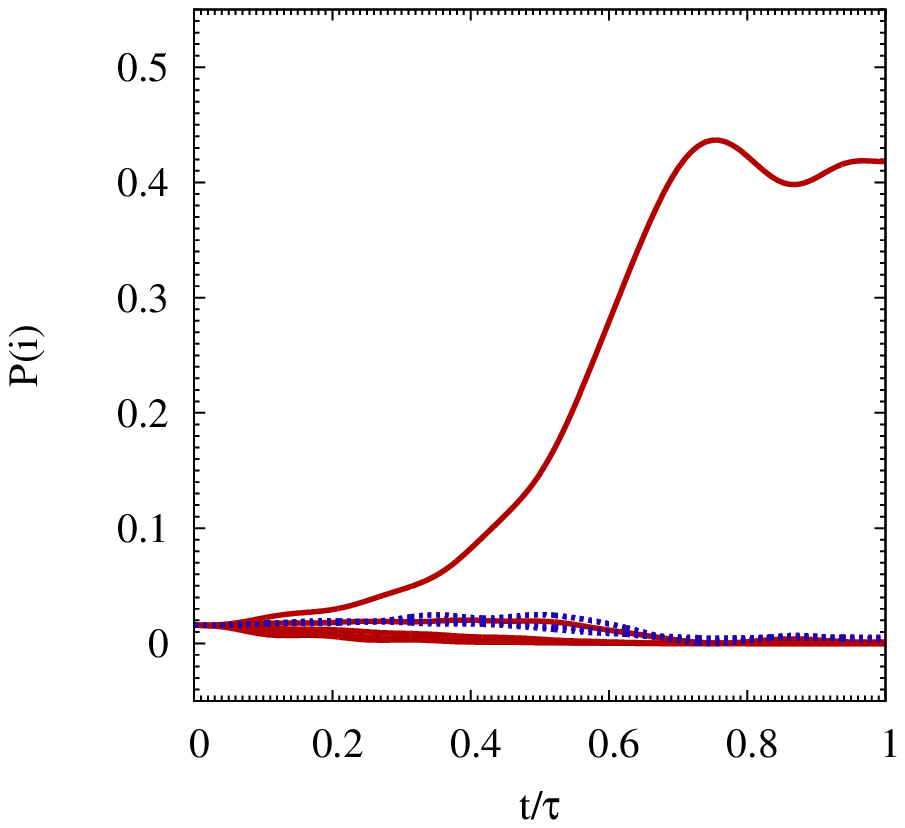}\\
({\rm e}) N_{\rm d} = 4, \tau = 1 &\hspace{1mm}& ({\rm f}) N_{\rm d} = 4, \tau = 10\\
  \end{array}$
  \caption{
  (Color online) Probability $P(i)$ for classical basis states $i$ in case of $N_{\rm d} = 1, 2$ and $4$ for $\tau = 1$ and $10$.
The curves have the same interpretations as in Fig.~\ref{graph:prob-dis-adia}.
  }
  \label{graph:Q-dyn-prob}
 \end{center}
\end{figure}

Next we consider the microscopic mechanism of this non-monotonic dynamics.
We calculate the real-time dynamics of the probability $P(i)$ of the classical bases $(i=(+,+,+),\cdots, (-,-,-))$ for $N_{\rm d}=1,2,$ and $4$ for $\tau = 1$ and $10$. The results are shown in Fig.~\ref{graph:Q-dyn-prob}.
There are $2^{N_{\rm d}}$ curves, but many of them are the same and overlap.
As $\tau$ increases, $P(i)$s approach the adiabatic ones shown in Fig.~\ref{graph:prob-dis-adia}.

If we consider probability $P_-$ to have $\sigma_1^z \sigma_2^z = -1$ which is the sum of $P(-,+,+)$,$P(-,-,+)$,$P(+,+,-)$,and $P(+,-,-)$, it  is zero in the adiabatic limit.
However, in the short time, for $\tau = 1$, $P_-$ remains nonzero.
For $\tau=10$, $P_-$ is close to zero. 

\begin{figure*}[t]
 \begin{center}
  $\begin{array}{ccccc}
   \includegraphics[trim=40 0 40 0,scale=0.5]{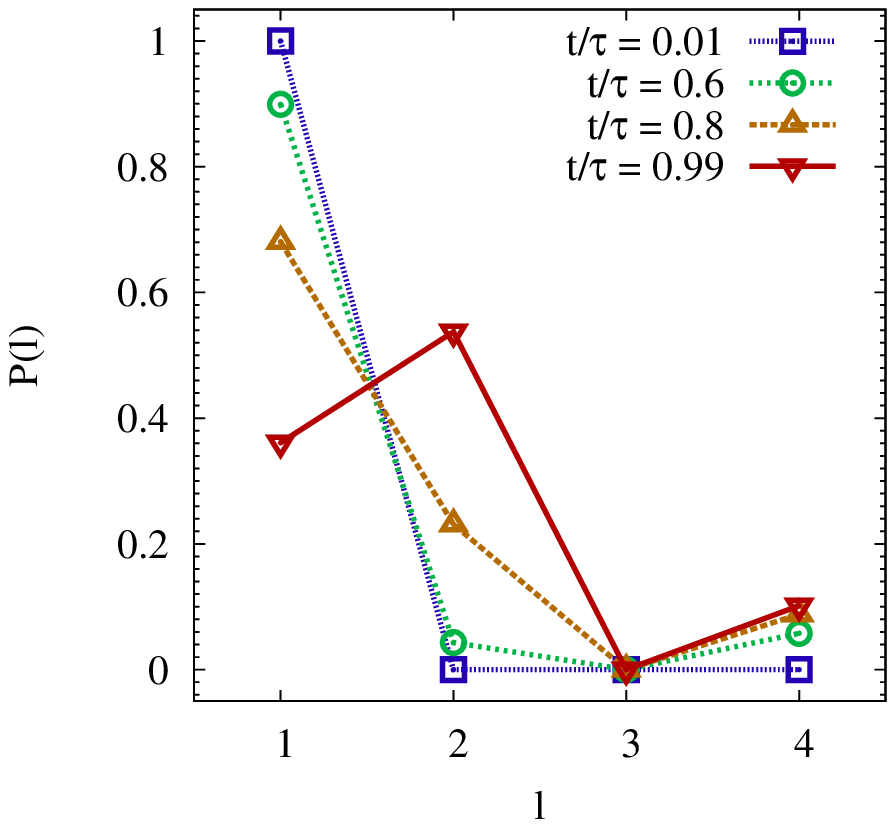}
    & \hspace{3mm} &
   \includegraphics[trim=40 0 40 0,scale=0.5]{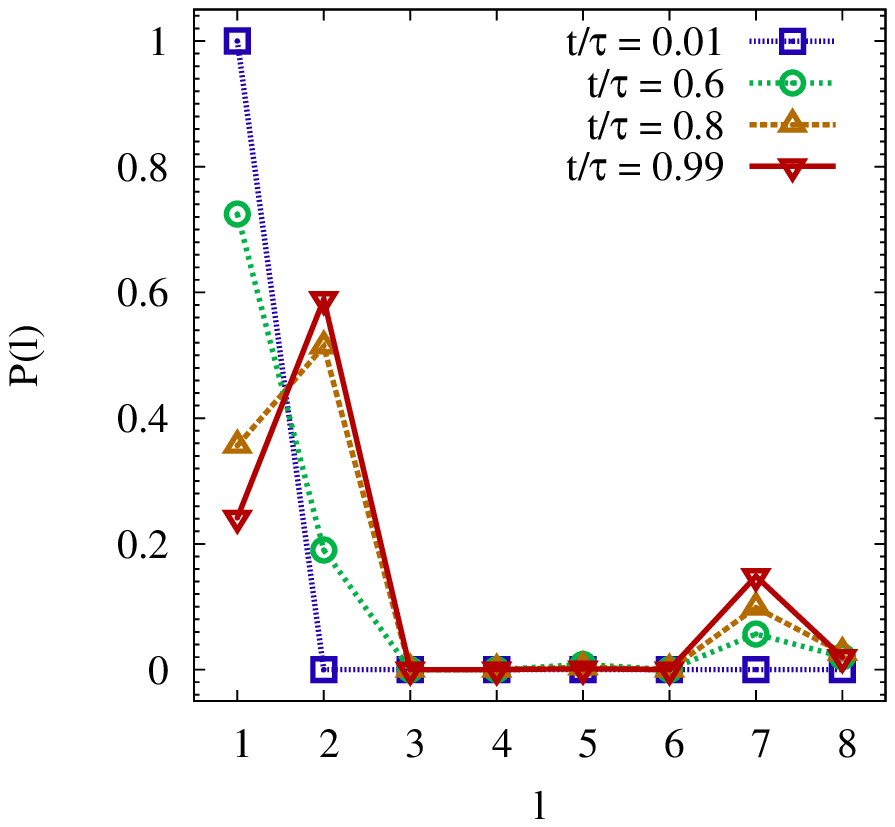}
    & \hspace{3mm} &
   \includegraphics[trim=40 0 40 0,scale=0.5]{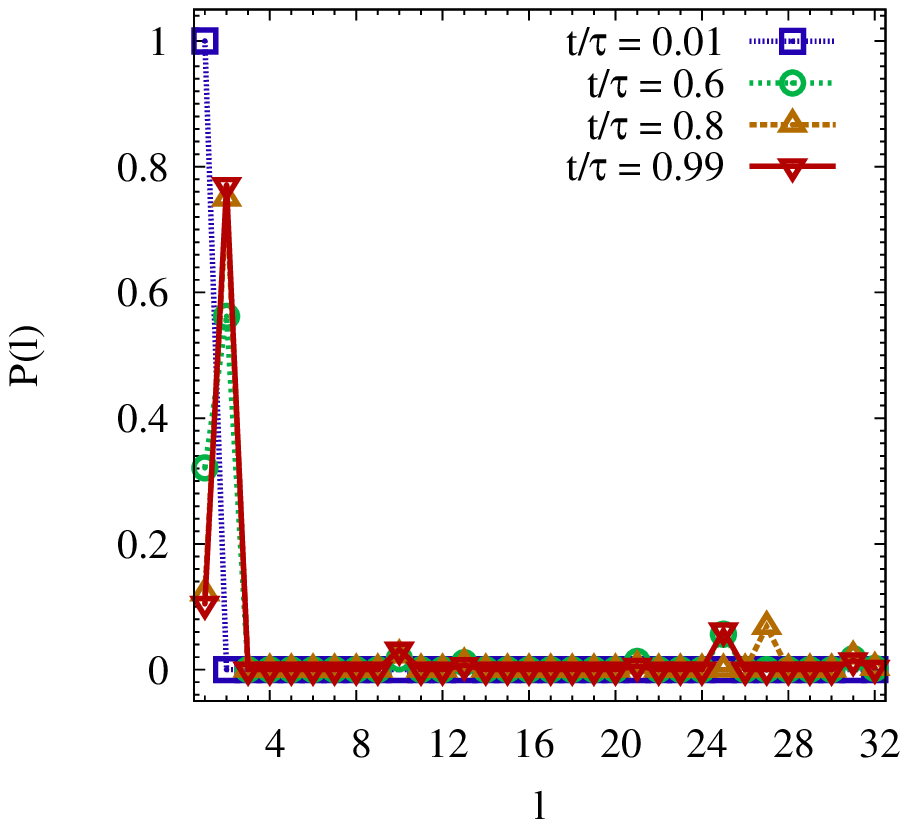}\\
    ({\rm a}) N_{\rm d} = 1, \tau = 1& \hspace{3mm} & ({\rm
     b}) N_{\rm d} = 2, \tau = 1& \hspace{3mm} & ({\rm c}) N_{\rm d} = 4, \tau = 1\\
       \includegraphics[trim=40 0 40 0,scale=0.5]{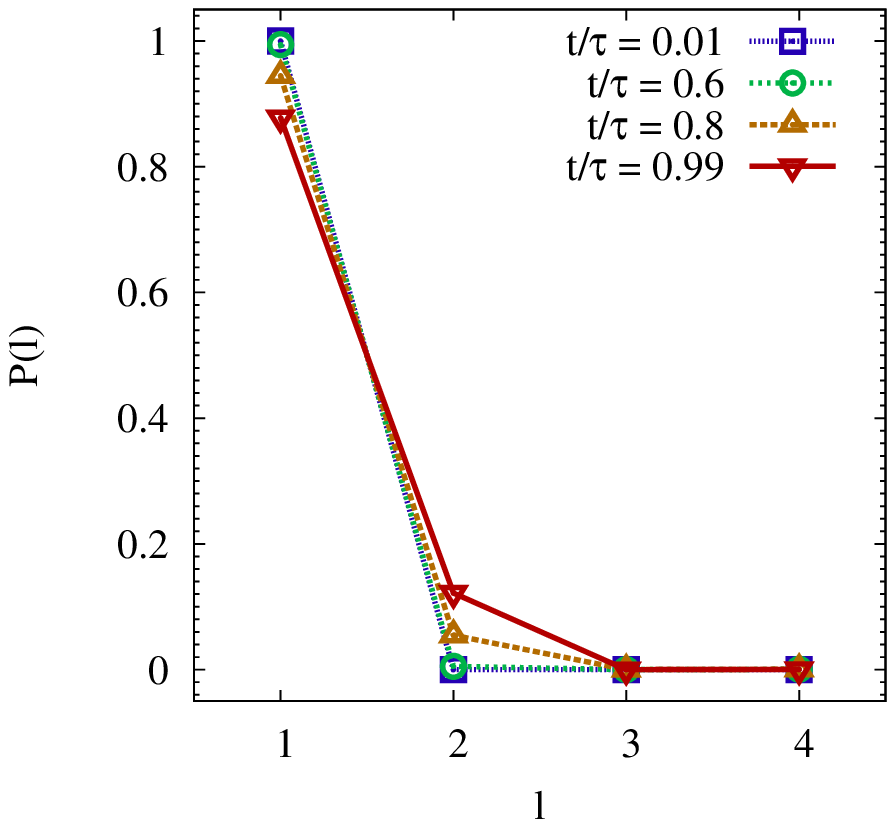}
    & \hspace{3mm} &
   \includegraphics[trim=40 0 40 0,scale=0.5]{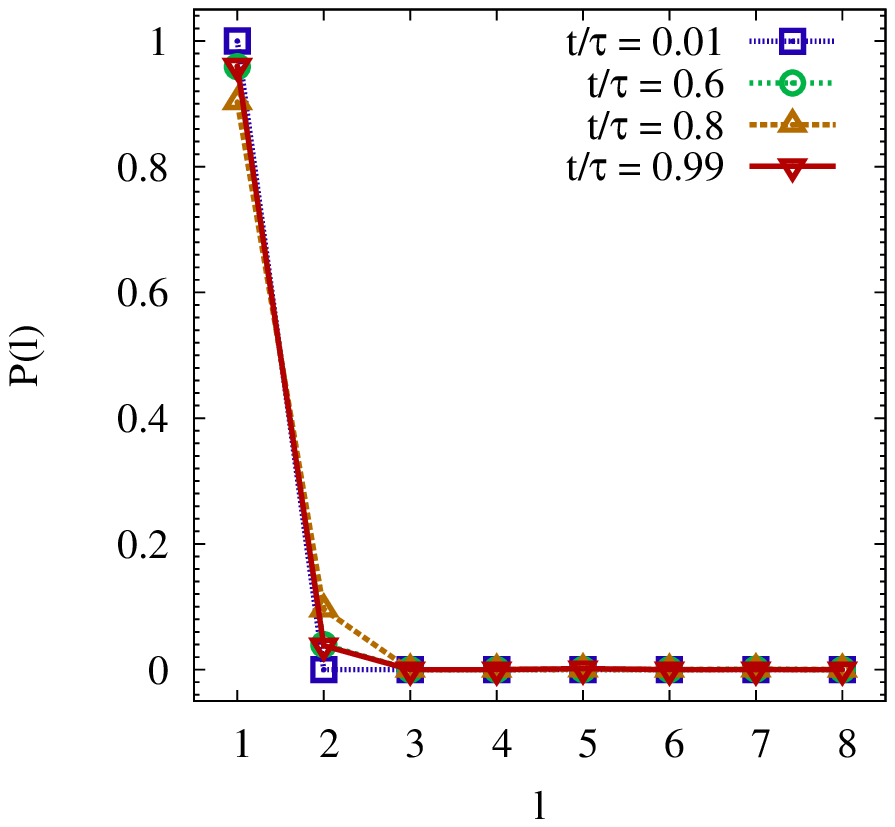}
    & \hspace{3mm} &
   \includegraphics[trim=40 0 40 0,scale=0.5]{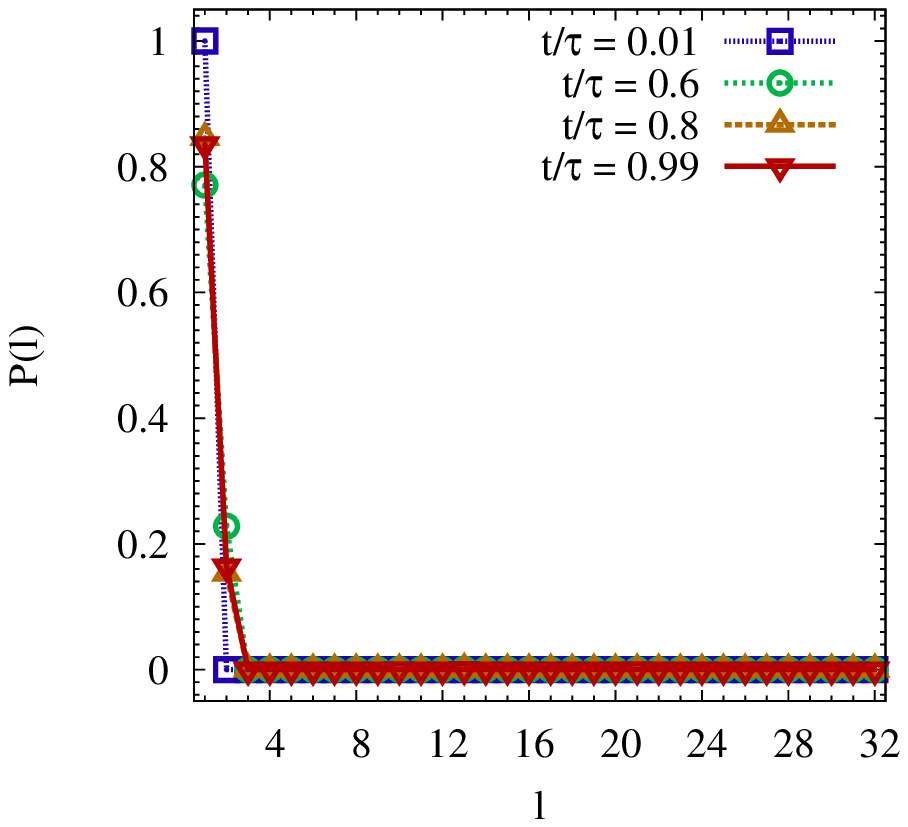}\\
    ({\rm d}) N_{\rm d} = 1, \tau = 10 & \hspace{3mm} & ({\rm
     e}) N_{\rm d} = 2, \tau = 10& \hspace{3mm} & ({\rm f}) N_{\rm d} = 4, \tau = 10
  \end{array}$
  \caption{
  (Color online) The probability $P(l)$ of eigenstates  $(l=1,2,\cdots,2^{N_{\rm d}-1})$.
Here , $l=1$ indicates the ground state.
  }
  \label{graph:Q-dyn-problevels}
 \end{center}

\end{figure*}

We also calculate the dynamic behavior of the probability of the eigenstate $\left\{ l \right\}$ at several times $t/\tau$. Here, the eigenstates are labeled in the order of eigenenergy ($l=1,2,\cdots,8$). Because the wavefunction after changing the Hamiltonian is symmetric, it is only necessary to consider the eigenenergy of symmetric wave functions. Figure \ref{graph:Q-dyn-problevels} shows the dynamical behavior of the probability at the state $\left\{ l \right\}$ in cases of $N_{\rm d}=1$, $2$, and $4$ for $\tau = 1$ and $10$.
For $\tau = 1$, the probability of the first excited level is larger than the probability of the ground state. On the other hand, for $\tau = 10$, the probability distribution decreases monotonically as a function of the eigenenergy.
As $\tau$ increases, the probability of the ground state approaches unity.

\section{Conclusion}

We studied a correlation function in a frustrated quantum spin system with a transverse field.
In the ground state, the system correlation function is non-monotonic as a function of the quantum parameter $\alpha$, as is the equilibrium value of the system correlation function as a function of temperature in the classical case.
We also considered the dynamics of the system correlation function of the system with a time-dependent quantum parameter.
When we increased the quantum parameter $\alpha$ fast, the wave function changed little from the initial wave function, and the dynamics is monotonic. 
As the sweeping speed of the quantum parameter $\alpha$ decreases, the dynamics of the system correlation function approaches the adiabatic limit, and the non-monotonic relaxation appears.
In the classical system, however, we found non-monotonic relaxation of the system correlation function when we decreased the temperature suddenly.
This fact shows the difference between the effects of the thermal fluctuation and the quantum fluctuations.
In frustrated systems, correlation function often behaves non-monotonic due to a peculiar density of states.
We expect that the mechanism which is discussed in this manuscript appears in real frustrated systems.

We thank Masaki Hirano, Hosho Katsura and Eric Vincent for fruitful discussions.
This work was partially supported by Research on Priority Areas 
``Physics of new quantum phases in superclean materials'' (Grant
No. 17071011) from MEXT and 
by the Next Generation Super Computer Project, Nanoscience Program from MEXT.
S.T. is partly supported by Grant-in-Aid for Young Scientists Start-up
(No.21840021) from JSPS.
The authors also thank the Supercomputer Center, Institute for Solid State Physics, University 
of Tokyo for the use of the facilities. 

\newpage 

\end{document}